\documentclass[american]{article}
\usepackage[T1]{fontenc}
\usepackage[latin1]{inputenc}
\usepackage{a4wide}
\usepackage{graphicx}
\usepackage{amssymb}

\makeatletter

 \newcommand{\lyxaddress}[1]{
   \par {\raggedright #1 
   \vspace{1.4em}
   \noindent\par}
 }

\usepackage{babel}
\makeatother
\begin{document}

\title{Evaluation of excitation energy and spin from light charged
particles multiplicities in heavy-ion collisions}

\author{J.C. Steckmeyer$^{1}$\footnote{Corresponding author, e-mail address :
Jean-Claude.Steckmeyer@lpccaen.in2p3.fr, tel : +33 2 31 45 29 66, fax : +33 2 31
45 25 49}$\ $,
Z. Sosin$^{2}$, K. Grotowski$^{2,3}$, 
P. Paw{\l}owski$^{3}$,\\
S. Aiello$^{4}$, A. Anzalone$^{5}$, M. Bini$^{6}$, B. Borderie$^{7}$,
R. Bougault$^{1}$, \\ G. Cardella$^{4}$, 
G. Casini$^{6}$, S. Cavallaro$^{5}$, J.L. Charvet$^{8}$,
R. Dayras$^{8}$, \\ E. De Filippo$^{4}$, D. Durand$^{1}$, 
S. Femin\`o$^{9}$, J.D. Frankland$^{10}$, E. Galichet$^{7,11}$, \\
M. Geraci$^{4}$, F. Giustolisi$^{5}$, P. Guazzoni$^{12}$,
M. Iacono Manno$^{5}$, G. Lanzalone$^{5}$, \\ G. Lanzan\`o$^{4}$,
N. Le Neindre$^{7}$, S. Lo Nigro$^{4}$, F. Lo Piano$^{5}$,
A. Olmi$^{6}$, \\ A. Pagano$^{4}$, M. Papa$^{5}$, M. Parlog$^{13}$,
G. Pasquali$^{6}$, S. Piantelli$^{6}$, \\
S. Pirrone$^{4}$, G. Politi$^{4}$, F. Porto$^{5}$, M.F. Rivet$^{7}$,
F. Rizzo$^{5}$, E. Rosato$^{14}$, \\
R. Roy $^{15}$, S. Sambataro$^{4}$\footnote{deceased}$\ $, M.L. Sperduto$^{4}$,
A.A. Stefanini$^{6}$, C. Sutera$^{4}$, \\ B. Tamain$^{1}$,
E. Vient$^{1}$, C. Volant$^{8}$, J.P. Wieleczko$^{10}$ and L. Zetta$^{12}$\\
(The INDRA and CHIMERA collaborations)}

\date{~}

\maketitle

\lyxaddress{
$^{1}$LPC, IN2P3-CNRS, ENSICAEN et Université de Caen,
14050 Caen-cedex 04, France. \\
$^{2}$Institute of Physics, Jagellonian University, Reymonta 4, 30-059
Kraków, Poland. \\
$^{3}$Institute of Nuclear Physics PAN, Radzikowskiego 152, 31-342
Kraków, Poland. \\
$^{4}$INFN Sezione di Catania and Dipartimento di Fisica e Astronomia,
Universit\`a di Catania, 95123 Catania, Italy. \\
$^{5}$LNS and Dipartimento di Fisica e Astronomia, Universit\`a di Catania,
95123 Catania, Italy. \\
$^{6}$INFN e Universit\`a di Firenze, 50125 Firenze, Italy. \\
$^{7}$Institut de Physique Nucléaire, IN2P3-CNRS, 91406 Orsay-cedex,
France. \\
$^{8}$DAPNIA/SPhN, CEA/Saclay, 91191 Gif-sur-Yvette-cedex, France. \\
$^{9}$INFN e Universit\`a di Messina, 98100 Messina, Italy. \\
$^{10}$GANIL, CEA et IN2P3-CNRS, BP 5027, 14076 Caen-cedex 05, France. \\
$^{11}$CNAM, Laboratoire des Sciences Nucléaires, 2 rue Conté, 75003 Paris,
France. \\
$^{12}$INFN e Universit\`a di Milano, 20133 Milano, Italy. \\
$^{13}$National Institute for Physics and Nuclear Engineering,
Bucharest-Magurele, Romania. \\
$^{14}$INFN e Dipartimento di Scienze Fisiche, Universit\`a di Napoli
{}``Federico II'', Napoli, Italy. \\
$^{15}$Laboratoire de Physique Nucléaire, Université Laval, Québec,
Canada.
}

\vspace{-17cm}

\emph{To be submitted to Physics Letters B}
\today

\vspace{17cm}

\begin{abstract}
A simple procedure for evaluating the excitation energy and the spin
transfer in heavy-ion dissipative collisions is proposed. It is based
on a prediction of the GEMINI evaporation code : for a nucleus with
a given excitation energy, the average number of emitted protons
decreases with increasing
spin, whereas the average number of alpha particles increases.
Using that procedure
for the reaction $^{107}$Ag+$^{58}$Ni at 52 MeV/nucleon, the excitation
energy and spin of quasi-projectiles have been evaluated. The results
obtained in this way have been compared with the predictions of
a model describing the primary dynamic stage of heavy-ion collisions.



\end{abstract}


%


In intermediate energy heavy-ion collisions at 
large impact parameters, a significant part of
the initial kinetic energy is converted into heat inside the
partners moving away from each other in the exit channel.
Nucleon exchange
as well as nucleon-nucleon (NN) collisions are responsible for the
dissipation of energy, creating two main excited fragments
namely the quasi-projectile
(QP) and the quasi-target (QT) \cite{Schroder}. Meanwhile, a fraction
of the initial orbital angular momentum is transferred into
intrinsic angular momentum (or spin) of the fragments. 

Decaying QP's and QT's are observed as sources of nucleons, light
charged particles (LCP's), intermediate mass fragments (IMF's) or
fission fragments for heavier systems, and $\gamma$-rays. Besides,
a third source of particles appears in the region between
the QP and QT fragments (see e.g. \cite{Pawlowski,thomas,diane}).
The contribution of such an intermediate velocity source (IVS) depends on
the initial angular momentum and masses of the colliding nuclei
in the entrance channel.

Production of excited nuclei with large spins and large excitation energies
is important to study the de-excitation properties of hot nuclear
matter. In particular, the role of angular momentum in the multifragmentation
process has been emphasized \cite{Botvina}. The nucleus spin is usually
evaluated from
the angular distribution of the emitted products. Measurements
using $\gamma$-rays \cite{Natowitz}, LCP's \cite{Babinet} and fission
fragments \cite{Dyer} have been performed mainly at low bombarding energies,
but scarce measurements exist in the intermediate energy range
\cite{Colin,jcs2}. In this Letter, we propose a simple procedure for evaluating
the excitation energy and spin transfer in heavy-ion dissipative collisions.

Binary dissipative collisions in the $^{107}$Ag+$^{58}$Ni reaction
have been studied at GANIL in inverse kinematics using a $^{107}$Ag beam at
52 MeV/nucleon
\cite{Steckmeyer}. For that purpose, the standard INDRA setup \cite{Pouthas}
was modified to detect QT nuclei recoiling with low kinetic energies of 
typically a few tens of MeV. Ten large area (5\ensuremath{×}5
cm$^{2}$) Si detectors, each of them having four vertical strips,
were used for the detection of QT nuclei in discrete angular
domains ranging from 3° to 87°. They were placed in the horizontal
plane of INDRA. The mass of the secondary QT (QT residue)
was deduced from energy and time-of-flight (TOF) measurements. As
in the 52 MeV/nucleon $^{107}$Ag+$^{58}$Ni reaction the QP
is expected to be emitted at very forward angles, the first ring of
the INDRA detector made of 12 phoswich plastic scintillators (2°<$\theta$<3°)
was replaced by the first wheel of the CHIMERA detector \cite{Aiello},
mounted at a distance of 4 m from the target in a dedicated vacuum
chamber. The CHIMERA detectors consisted of 2 rings of 16 Si detector
- CsI scintillator telescopes, allowing for a better identification
of the atomic number of the secondary QP (QP residue) in the angular
range 1°<$\theta$<3°. Event acquisition was triggered by the detection
of a charged product in one of the 10 TOF-Si detectors. The binary
events were selected by requiring that the
mass of that nucleus, assumed to be the QT residue, be larger than
20 u. Such conditions selected peripheral and
semi-peripheral collisions. Only complete
events were kept in the analysis: events with total detected charge
and momentum larger than 90 \% of the incident charge and momentum,
respectively. In the following we will concentrate in studying
the properties of the primary QP.
It has been reconstructed with particles emitted
in its forward hemisphere, their contribution being counted twice.
In doing so, most of particles emitted from other sources
are assumed being not accounted for. For the data set considered
in this work, the reconstructed average primary QP charge is 47$\pm$2. 

Let $E^{*}/A$ and $J$ denote the excitation energy per nucleon 
and spin of the primary QP, respectively.
For the purpose of the $E^{*}/A$ and $J$ evaluation procedure (in
this paper referred to as the \emph{procedure}), the Monte-Carlo code
GEMINI \cite{Charity} was used to simulate the de-excitation stage
of the primary QP. Standard prescriptions were used with a temperature
dependent level density parameter \cite{lestone}.
In the framework of that model and for a given
nucleus, the knowledge of the average  proton and
alpha particle multiplicities, $\overline{M}_{p}$
and $\overline{M}_{\alpha}$,
makes possible the estimation of the excitation energy and spin of
that nucleus: a given ($\overline{M}_{p},
\overline{M}_{\alpha}$) couple corresponds to
a unique ($E^{*}/A$, $J$ ) couple. Such a determination is valid
as long as the multiplicities predicted by GEMINI are not too far
from the experimental values, and the considered particles really
come from the nucleus under study. Similar attempts were made
to determine the fragment spin by looking at the ratio of H to He isotopes
\cite{casini}.

The multiplicity of IMF's emitted by the primary QP
is low compared to that of LCP's \cite{Steckmeyer} :
in the most dissipative collisions studied here, $\overline{M}_{p}$,
$\overline{M}_{\alpha}$
and $\overline{M}_{IMF}$ are 1.71, 1.19 and 0.16, respectively,
in the forward hemisphere. Consequently, the GEMINI
calculations will be performed allowing only for neutron and LCP evaporation
(hydrogen and helium nuclei) and no IMF emission.
Since among the LCPs,
the $^{1}$H and $^{4}$He isotopes prevail, further on we will mention
mainly protons and alpha particles. 

\begin{figure}
\begin{center}\includegraphics[%
  width=8cm,
  height=8cm,
  keepaspectratio]{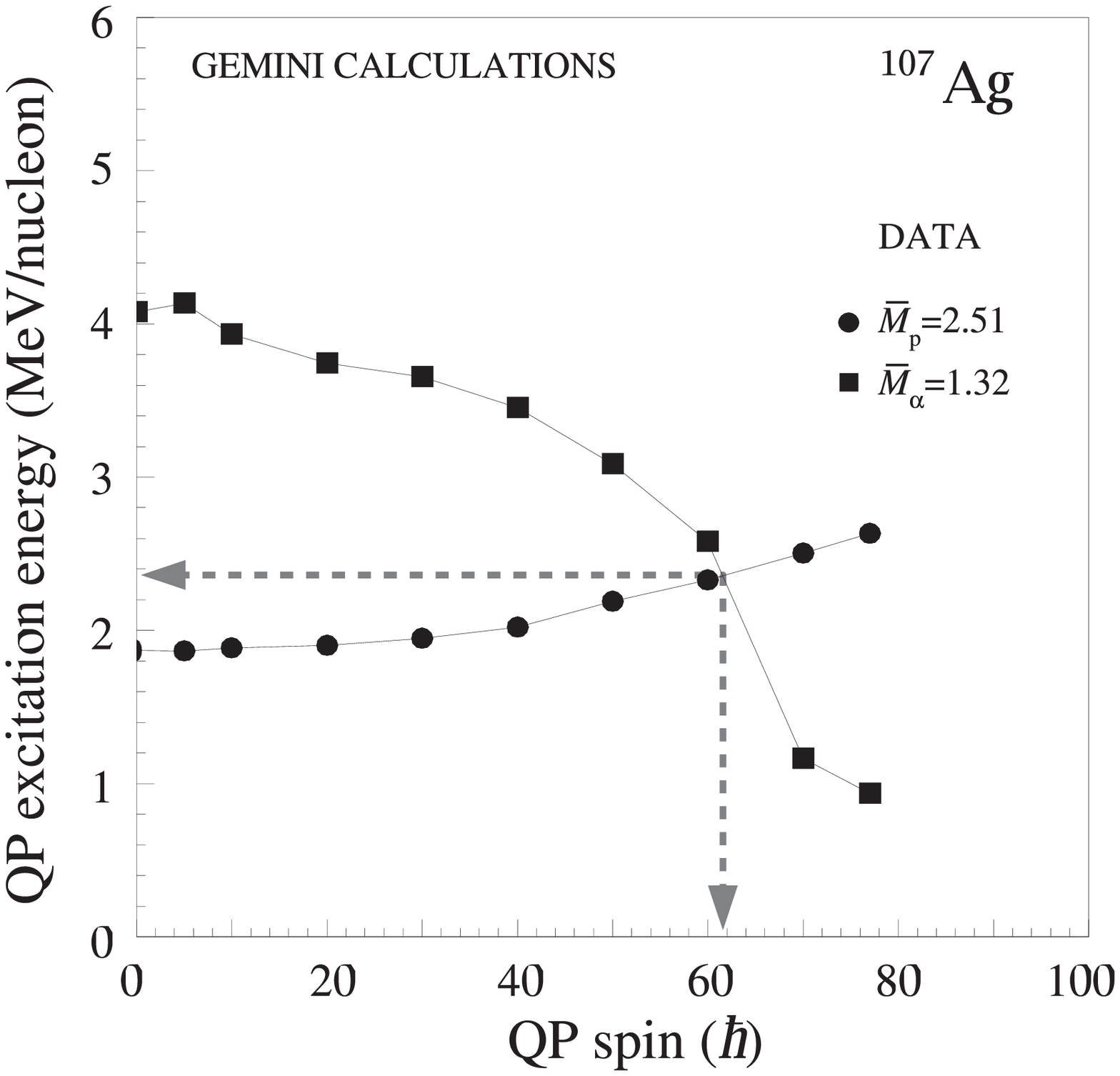}\end{center}

\caption{\label{Procedure}$\overline{M}_{p}=const$ and
$\overline{M}_{\alpha}=const$ curves in
the $E^{*}/A$ excitation energy vs. $J$ spin plane as generated by the
GEMINI calculations.}
\end{figure}

As predicted by GEMINI, both $\overline{M}_{p}$
and $\overline{M}_{\alpha}$ increase with increasing
excitation energy. However, for a given excitation energy, 
$\overline{M}_{p}$
decreases with increasing spin, whereas $\overline{M}_{\alpha}$
increases, i.e. the higher the spin, the
heavier the mass of the evaporated particle. Indeed, the best way for a
fast rotating nucleus to release its spin is to emit heavy ejectiles
which carry away high angular momentum.
Such an effect of the nucleus spin on the
probability of particle evaporation has also been seen in \cite{dipietro}.
This opposite behavior of $\overline{M}_{p}$ as compared to 
$\overline{M}_{\alpha}$ has been used
in the \emph{procedure} to extract the correlation between
the excitation energy and the spin of the primary QP.
GEMINI predictions can be used to draw the $E^{*}/A$
vs. $J$ curve associated with a constant $\overline{M}_{p}$,
whatever the multiplicity of other LCP's.
The same can be done for a constant $\overline{M}_{\alpha}$
multiplicity. As shown in Fig. \ref{Procedure},
these two curves exhibit an opposite trend, their intersection point
defining the $E^{*}/A$ and $J$ values. Using this correspondence
one can associate $(E^{*}/A$ , $J$) values to each ($\overline{M}_{p}$,
$\overline{M}_{\alpha}$) pair measured in the experiment. 

In order to find out the evolution of $E^{*}/A$ and $J$ of the
primary QP
as a function of the dissipation, the events were sorted according to
the QP residue velocity parallel to the beam,
determined from the energy and the mass estimated from a fit
performed on the stability valley. The associated
velocity spectrum has been divided in twelve bins of equal width, going from
0.88 to 0.99 times the projectile velocity $V_P$.
For these bins the $\overline{M}_{p}$ and $\overline{M}_{\alpha}$
multiplicities have been measured and the $E^{*}/A$ and $J$
values have been evaluated using the
\emph{procedure} shown in Fig. \ref{Procedure}.
GEMINI calculations were performed
for a $^{107}$Ag nucleus because the reconstructed primary QP has a mean
atomic number of 47, as mentioned previously.
No attempt was made to account for the slight
variation of the mean value of the atomic number as a function of the
QP velocity nor for
the width of its distribution. Such an analysis is beyond the scope
of this work, mainly focused on the gross properties of the QP.

\begin{figure}
\begin{center}\includegraphics[%
  width=12cm,
  height=8cm,
  keepaspectratio]{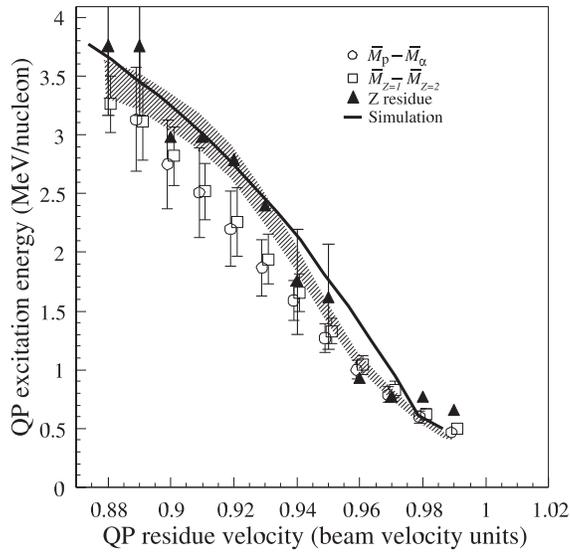}\end{center}

\caption{\label{E_vs_Vp}
As a function of the QP residue velocity, the
excitation energy of the primary QP
deduced from the charge of the detected QP residue (black triangles),
from the calorimetry method (hatched band) and from the
($\overline{M}_p, \overline{M}_{\alpha}$) and 
($\overline{M}_{Z=1}, \overline{M}_{Z=2}$) experimental data
(open circles and squares).}
\end{figure}

In the GEMINI calculations, at any given excitation energy,
the charge of the QP residue $Z^{res}_{QP}$ is found to be practically
independent of the spin of the primary QP, 
suggesting that one can determine the excitation energy of the primary QP
from the measurement of $Z^{res}_{QP}$ \cite{Steckmeyer}. 
Such a determination is
presented as triangles in Fig. \ref{E_vs_Vp}. The error bars displayed at QP
residue velocities of .88-.89 $V_p$ and .94-.95 $V_P$ are estimates of the
uncertainty on the $E^{*}/A$ linked to an uncertainty of 2 charge units on
the charge of the primary QP. The excitation
energy can also be deduced from the experiment by using the calorimetry
method \cite{Peter} : the excitation energy of the primary QP is deduced from
the kinetic energies of the emitted particles. In this evaluation, 
the neutron emission has
been estimated from the difference between the mass of the primary QP,
assumed to have the same neutron to proton ratio as
the projectile, and the masses of all de-excitation products \cite{jcs2}.
As one can see, the results of the calorimetry method,
displayed as the hatched band in Fig. \ref{E_vs_Vp}, agree well with
the results obtained from the atomic number of the QP residue.
Also a reasonable agreement with the values deduced from the
LCP multiplicities is to be noted in the same figure
(open circles deduced from the $\overline{M}_{p}$ and $\overline{M}_{\alpha}$
multiplicities and open squares from $\overline{M}_{Z=1}$ and 
$\overline{M}_{Z=2}$ multiplicities).
However, the results obtained from the
$Z^{res}_{QP}$ charge are slightly higher, particularly above $\approx$
2 MeV/nucleon.
The experimental charge $Z^{res}_{QP}$ accounts for emission of LCP's
as well as IMF's. From the experimental value $Z^{res}_{QP}$
and using the $E^{*}/A$ - $Z^{res}_{QP}$ correlation as
calculated with GEMINI (taking into account only LCP emission), we deduce,
in some sense, the excitation energy of a primary QP having emitted both LCP's
and IMF's; a fraction of the experimental charge loss
$\Delta Z = Z^{prim}_{QP} - Z^{res}_{QP}$ being associated with IMF emission.
At variance, the $E^{*}/A$ deduced from the LCP multiplicity measurements
reflects the $E^{*}/A$ of a primary QP having only emitted H and He isotopes,
the excitation energy of which is slightly
lower than the one of the same nucleus having
emitted the same numbers of H and He isotopes plus a few IMF's.

\begin{figure}
\begin{center}\includegraphics[%
  width=8cm,
  keepaspectratio]{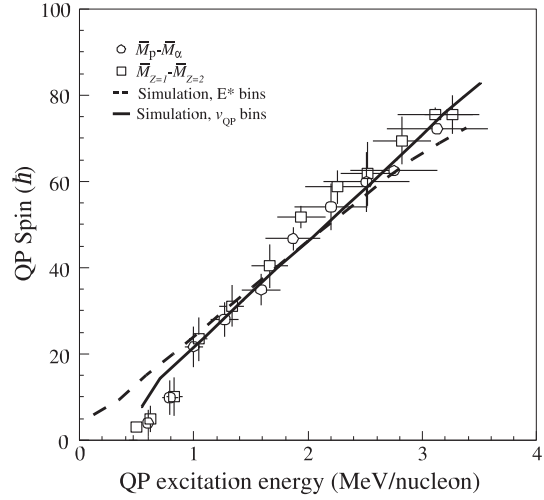}\end{center}

\caption{\label{Spin_vs_E}Spin of the primary quasi-projectile vs. its
excitation
energy. For open circles the proton and alpha particle
multiplicities were used, and
for open squares the multiplicities of $Z=1$ and $Z=2$ particles.
Solid and broken lines denote model simulations. The solid line is the
excitation energy - spin correlation obtained when data are sorted as a 
function of the QP residue velocity and the
broken line when data are sorted as a function of the 
excitation energy. }
\end{figure}

In Fig. \ref{Spin_vs_E},
the open circles show the $E^{*}/A - J$ correlation extracted from the
($\overline{M}_{p}$ , $\overline{M}_{\alpha}$) experimental
data and the open squares the same correlation extracted from
the multiplicities of $Z=1$ and $Z=2$ particles, using the
same \emph{procedure}. As
can be seen, the spin increases linearly with increasing excitation
energy (except below $\approx$ 1 MeV/nucleon) and 
reaches values up to 70-80 $\hbar$ at excitation
energies of 3-3.5 MeV/nucleon. This maximum value is of the order of magnitude
of the angular momentum at which the fission
barrier of a $^{107}$Ag nucleus vanishes as predicted
by the Fast Rotating Liquid Drop Model \cite{Sierk}. 

The $E^{*}/A - J$ thus extracted is compared to the prediction
of the Sosin model \cite{Sosin}, referred herein as the \emph{model}.
This \emph{model} does not assume full thermalization of the system
of colliding heavy ions. Instead, individual fragments (clusters)
are thermally equilibrated. Therefore the state densities can be applied
for determining the configuration probabilities. 

In the \emph{model} a two-stage reaction scenario is assumed for a
heavy ion collision which finally creates \char`\"{}hot sources\char`\"{}
of particles. In the first stage, some of the nucleons become active
reaction participants either via mean field effects or via NN interactions.
In the second stage, these active nucleons may undergo coalescence,
be transferred to the target remnant, to the projectile remnant, or
to some clusters created earlier. Alternatively, they may escape to
the continuum. The two stages of the reaction scenario do not mean
a time sequence, but only a factorization of the total probability
of particular configurations of particle systems. 

In the mean field mechanism, one of the nucleons of the projectile
nucleus or target nucleus becomes an active participant when running
across a potential window which opens in the region between the colliding
heavy ions.
Both size and time during which the window remains open
depend on the proximity
and relative velocity of the heavy ions along their classical Coulomb
trajectories. These trajectories are calculated for the interaction
potentials (Coulomb and nuclear). 

In the NN mechanism the two nucleons, one from the projectile and
the other one from the target, collide in their overlap zone,
where - for collisions at high energy and/or at large impact parameters
- the Pauli principle becomes less restrictive. The nucleons of such
a pair become reaction participants. The probability of a NN collision
depends on the cross-section of the NN interaction, the convolution
of the projectile and the target nucleus densities in the overlap
region, as well as on the available momentum space. 

%

The nucleon transfer process is executed in a series of steps. A detailed
description of this process is given in \cite{Sosin}. The cluster
coalescence process is possible when two clusters running along their
trajectories are being trapped in a potential well. 

A separate problem concerns the distribution of the excitation energy
available at each step of the nucleon transfer process. The total energy
of the system
is of course conserved along the chain of transfers, but the excitation
energies of individual subsystems vary according to the particular reaction
$Q$ value. For a given step $k$ of nucleon transfer, the summation
of the ground state and kinetic energies of fragments with their interaction
potentials calculated for the exit reaction channel provides a value
of the total energy corresponding to the internal degrees of freedom
(excitation energy) of the system after step $k$. After subtracting
the total excitation energy of the step $k-1$, one obtains the corresponding
reaction $Q$ value which is divided among all
involved subsystems of mass $A>4$, with probability proportional
to the corresponding densities of states. 

The orbital angular momenta and intrinsic spins of the
primary reaction products are
calculated from the angular momenta of all nucleons of the system.
The angular momenta of the nucleons are calculated assuming
a Fermi gas distribution and that the
initial locations depend upon the mean field and the NN interaction
mechanism. 

The \emph{model} described above, coupled to the GEMINI evaporation
code, has been successfully applied for description of reactions in
medium heavy systems \cite{Planeta,Buta}. 

For the reaction studied in this paper, the QP mass predicted by the
\emph{model} is $A=107\pm4$ (peripheral collisions); the contribution
of the IVS protons and alphas emission in the forward hemisphere
in the reference frame of the primary QP varies
from 5 to 10\% (depending on the impact parameter). Note, that these
\emph{model} predictions justify the assumptions made
in the reconstruction of the primary QP and the choice of a $^{107}$Ag nucleus
in the GEMINI calculations for application of the \emph{procedure}. However,
energy is dissipated by the IVS. In the most dissipative collisions
studied in this work,
the model predictions lead to
an excitation energy of the primary QP of $\approx$ 365 MeV, whereas the energy
associated with the IVS emission reaches $\approx$ 445 MeV. These results are
quite in agreement with a recent analysis performed on the $^{93}$Nb+$^{93}$Nb
reaction at 38 MeV/nucleon \cite{mangiarotti}. Looking in more details to the
calculations, it appears that on average (20$\pm$7) nucleons originate from the
IVS, 60\% of them coming from the QT nucleus nearly two times
lighter than the QP one. This explains in some way the weak evolution of the
QP mass as a function of the dissipation.

In the simulations presented below, we use predictions for the primary
dynamic reaction stage without de-excitation of the primary fragments.
We assume that
the QP has reached thermal equilibrium (see e.g. \cite{Planeta})
and that the evaporation of particles does not change its parallel velocity,
due to the forward-backward symmetry of the angular distribution of the
emitted LCP in the QP frame.
In consequence, the primary QP velocity should be, on average, equal
to the QP residue velocity.

Predictions of the \emph{model} for the $E^{*}/A - J$ correlation
of the primary QP are displayed as lines in Figs.
\ref{E_vs_Vp} and \ref{Spin_vs_E}. The solid line presents the \emph{model}
calculations when the events are sorted as a function of the QP residue
velocity and
the broken line (Fig. \ref{Spin_vs_E}) when data are sorted
as a function of the excitation energy, disregarding the QP residue velocity.
The \emph{model} predictions are in better agreement with the
excitation energies (Fig. \ref{E_vs_Vp}) obtained from the
atomic number of the QP residue
and from the calorimetry method, than with the ones obtained from average
LCP multiplicities.
It is likely due to the fact, as stated before, that
the IMF's are implicitly accounted for in the estimation of
the excitation energy. For the QP spin values,
the agreement between the \emph{model} and the \emph{procedure} predictions
is good in a broad range of excitation energies (Fig. \ref{Spin_vs_E}).
A noticeable discrepancy is observed only below 1 MeV/nucleon. It
should be pointed out that the \emph{model} assumes zero values for
the projectile and target intrinsic spins, and therefore the QP spin
should vanish in the zero-excitation-energy limit. It is really the
case for the \emph{model} (broken line), where calculations were
sorted as a function of the primary QP excitation energy.
Alternatively, when the calculations are sorted as a function of the
QP residue velocity bins, the QP spin
vanishes at about 0.5 MeV/nucleon for both the \emph{model}
and the \emph{procedure}. 
Such an effect could be explained by fluctuations and correlations between
the spin and the velocity generated by the reaction mechanism. 

In this work, the excitation energy of quasi-projectiles produced
in the $^{107}$Ag+$^{58}$Ni reaction at 52 MeV/nucleon has been
estimated in two distinct manners : from the calorimetry method and
from comparisons with GEMINI calculations, using either LCP
multiplicities or QP residue charge. The dynamical
Sosin model reproduces these results and
the overall agreement gives confidence in the evaluation of the
primary QP excitation energy.
The $E^{*}/A - J$ correlation is also well reproduced by the calculations,
indicating that the \emph{procedure}
is a powerful tool to determine 
the excitation energy and spin of excited nuclei
from LCP experimental multiplicities.
The agreement between the calorimetry method and the Sosin model
implies that the contribution of the IVS particles is low in the
forward hemisphere in the QP frame, otherwise the LCP multiplicities
should be higher and the excitation energy too. Dynamical calculations
presented in this paper
reproduce the features of the primary QP source, i.e. charge, excitation
energy and spin. By looking at the de-excitation step and cluster
formation and comparing with the data, it should be possible to evaluate
precisely the characteristics and properties of the IVS contribution.
It is also important to explain contribution of fluctuations and correlations
in the reaction picture. These are the goals of a forthcoming paper.

\textbf{Acknowledgements}

We are grateful to R. Bassini, C. Boiano, C. Cal\'i, M. D'Andrea, F. Fichera,
N. Guidice, N. Guardone, D. Nicotra, B. Raine, C. Rapicavoli, G. Rizza,
J. Ropert, J. Tillier, M. Tripon, S. Urso, J.L. Vignet and G. Wittwer for
their invaluable help in performing both mechanical, electronic and
acquisition coupling of the first ring of CHIMERA with INDRA.

\end{document}